\newcommand{\DocVersion}{\footnotesize \it Accepted for publication in the ApJ Letters SOFIA Special Issue, 15~Feb.~2012}

\documentclass[11pt,preprint]{aastex}

\slugcomment{\DocVersion}

\shortauthors{Shuping, et~al.}
\shorttitle{SOFIA/FORCAST Observations of the Orion Nebula}

\newcommand{\msun}{M$_{\sun}$}
\newcommand{\lsun}{L$_{\sun}$}

\newcommand{\thetaoned}{\object[* tet01 Ori D]{$\theta^{1}$~D~Ori}}
\newcommand{\thetaonec}{\object[* tet01 Ori C]{$\theta^{1}$~C~Ori}}

\newcommand{\lvone}{\object[OW94 168-326E]{LV1}}
\newcommand{\ircnine}{\object[2MASS J05351380-0521596]{IRc~9}}

\newcommand{\halpha}{H$\alpha$}

\begin{document}

\title{First Science Observations with SOFIA/FORCAST:  6 to 37~\micron\ Imaging of the Central Orion Nebula}

\author{R. Y. Shuping\altaffilmark{1}}
\affil{
Space Science Institute \\ 
4750 Walnut St.  \\
Suite 205  \\
Boulder, Colorado   \\
80301}
\email{rshuping@spacescience.org}

\and

\author{Mark R. Morris}
\affil{Department of Physics and Astronomy \\
University of California \\
Los Angeles, CA 90095-1547, USA}

\and

\author{Terry L. Herter, Joseph D. Adams, G. E. Gull, J. Schoenwald, \& C. P. Henderson}
\affil{Center for Radiophysics and Space Research \\
Cornell University \\
208 Space Sciences Building \\
 Ithaca, NY 14853 USA}

\and

\author{E. E. Becklin\altaffilmark{2}, James M. De Buizer, William D. Vacca, and Hans Zinnecker}
\affil{SOFIA-USRA \\
NASA Ames Research Center \\
MS N211-3 \\
Moffett Field, CA 94035, USA}

\and

\author{S. Thomas Megeath}
\affil{Dept. of Physics \& Astronomy \\
University of Toledo \\
2801 W. Bancroft St. \\
Toledo, OH 43606, USA}

\altaffiltext{1}{SOFIA-USRA,
NASA Ames Research Center,
MS N211-3,
Moffett Field, CA 94035, USA}
\altaffiltext{2}{Department of Physics and Astronomy,
University of California,
Los Angeles, CA 90095-1547, USA}

\begin{abstract}

We present new mid-infrared images of the central region of the Orion Nebula using the newly commissioned SOFIA airborne telescope and its 5 -- 40~\micron\ camera FORCAST.  The 37.1~\micron\ images represent the highest resolution observations ($\lesssim$~4\arcsec) ever obtained of this region at these wavelengths. After BN/KL (which is described in a separate letter in this issue), the dominant source at all wavelengths except 37.1~\micron\ is the Ney-Allen Nebula, a crescent-shaped extended source associated with \thetaoned.  The morphology of the Ney-Allen nebula in our images is consistent with the interpretation that it is ambient dust swept up by the stellar wind from \thetaoned, as suggested by \citet{Smith:2005}.  Our observations also reveal emission from two ``proplyds'' (proto-planetary disks), and a few embedded young stellar objects  (YSOs; \ircnine, and OMC1S IRS1, 2, and 10).  The spectral energy distribution for \ircnine\ is presented and fitted with  standard YSO models from \citet{Robitaille:2007} to constrain the total luminosity, disk size, and envelope size. The diffuse, nebular emission we observe at all FORCAST wavelengths is most likely from the background photodissociation region (PDR) and shows structure that coincides roughly with \halpha\ and [\ion{N}{2}] emission.  We conclude that the spatial variations in the diffuse emission are likely due to undulations in the surface of the background PDR.  

\end{abstract}



\section{Introduction}
\label{Intro}

Because of its relative proximity and its favorable location on the near side of its parent molecular cloud, the Orion Nebula (\object{M42}, \object{NGC1976}) has long been studied as the archetype of massive star formation.  Its infrared characteristics have been key to our understanding of the flow of energy and the role of dust in such a region.  Extended thermal emission from warm dust in the vicinity of the Trapezium OB star cluster that powers the Orion Nebula was discovered at 11.6 and 20~\micron\ by \citet{Ney:1969}.  More detailed observations of this ``Ney-Allen Nebula'' were subsequently carried out between 8 and 13~\micron\ by \citet{Gehrz:1975}, who showed that the emission peaks on \thetaoned\ (\object{HD37023}, B0.5V) and argued that the primary heating mechanism for the grains in this region is resonantly trapped Lyman-$\alpha$ radiation.  Using somewhat higher spatial resolution observations (3\arcsec), \citet{Hayward:1994} observed the Orion Nebula at multiple wavelengths from 7.6 to 24.3~\micron, and used the spectral energy distributions (SEDs) on representative sightlines to confirm an earlier claim by \citet{Forrest:1976} of the presence of a second, cooler, extended emission component.  Hayward showed that the cool component (110~K {\it vs.} 300~K for the warm component) peaks to the West of \thetaoned, and concluded that it arises from the back wall of the \ion{H}{2} region blister and is primarily heated by radiation from \thetaonec\ (\object{HD37022}, O4-6pv).  An overview of the geometry of the Orion Nebula and the ``Main Ionization Front'' (MIF) is given by \citet{ODell:2001}.  

Recent high-resolution mid-IR observations of the Orion Nebula have revealed a wealth of new detail~\citep{Smith:2005,Robberto:2005,Kassis:2006}.  The warm dust component is resolved largely into limb-brightened dust arcs around proplyds, and the brightest dust arc\footnote{
Usage of the term ``Ney-Allen Nebula'' has evolved to now refer uniquely to this bright feature, which lies at the peak of the extended structure discovered by \citet{Ney:1969}.}---that around \thetaoned---is either a consequence of that star being located close to the background cloud~\citep{Smith:2005} or is a photoevaporated circumstellar disk~\citep{Robberto:2005}.  The remaining, mostly cool dust emission emerges as a complex of streaks, gaps, arcs, and protostellar jets. 

In this letter we report on a study of the Orion Nebula with the Stratospheric Observatory for Infrared Astronomy (SOFIA) using the mid-IR facility camera FORCAST.   In this study,  we extend the relatively high-resolution view of the diffuse emission in this region to longer wavelengths than have previously been reported, and focus on the \object{Ney-Allen Nebula} and the proplyds.  Our report also includes brief descriptions of emission from the infrared source \ircnine\ and the cluster of young stellar objects in OMC1 South.  Results for the active star formation region surrounding the Becklin-Neugebauer source and the Kleinmann-Low Nebula are reported separately in this issue~\citep{DeBuizer:2012}.

\section{Observations and Analysis}
\label{sect:obs}

The {\it SOFIA} observations of Orion were performed on the nights of 30 November and 3 December 2010 using the FORCAST instrument~\citep{Adams:2010}. FORCAST is a dual-array mid-infrared camera capable of taking images with two filters simultaneously. The short wavelength camera utilizes a 256$\times$256 pixel Si:As array optimized for 5--25~\micron, while the long wavelength camera utilizes a 256$\times$256 pixel Si:Sb array optimized for 25-40~\micron ~\citep{Herter:2012}. After correction for focal plane distortion, FORCAST effectively samples at 0.768$\arcsec$ per pixel, which yields an approximately 3.4$\arcmin$$\times$3.2$\arcmin$ instantaneous field-of-view.

Observations were obtained in the 6.6, 7.7, 19.7, 31.5, and 37.1~\micron\ filters at an aircraft altitude of 43000 ft. The chopping secondary of {\it SOFIA} was driven at 2~Hz, with a chop throw of 7$\arcmin$. Such large throws were necessary to assure that the sky reference beam was completely off the extended Orion nebula complex and sampling ``empty'' sky areas for background subtraction. For the same reasons the telescope was nodded 10$\arcmin$ away from the Orion nebula. With this setup, Orion  was sampled in only one of the four chop-nod beams, and therefore the final effective on-source exposure times for the observations were 150~s at 6.6 and 7.7~\micron, 450~s at 19.7~\micron, 300~s at 31.5~\micron, and 420~s at 37.1~\micron.

Because all FORCAST data were taken in the dichroic mode, we were able to determine relative astrometry using images from the two wavelengths that were observed simultaneously. Data were taken with the following filter pairs: 19.7~\micron/37.1~\micron, 19.7~\micron/31.1~\micron, 6.6~\micron/37.1~\micron, and 7.7~\micron/37.1~\micron. Therefore the relative astrometry of all filters was determined precisely by bootstrapping from the 19.7~\micron/37.1~\micron\ pair. We estimate that the relative astrometry between filters is known to better than 0.5 pixels ($\sim$0.38$\arcsec$).  Absolute astrometry for each image was fixed by assuming that the peak of the BN source at 19.7~\micron\ is the same as the peak at 7~mm given by \citet{Rodriguez:2005},  R.A.(J2000)=05h35m14.119s, Dec(J2000)=-05d22m22.785s.

\begin{figure}
\epsscale{1.0}
\plotone{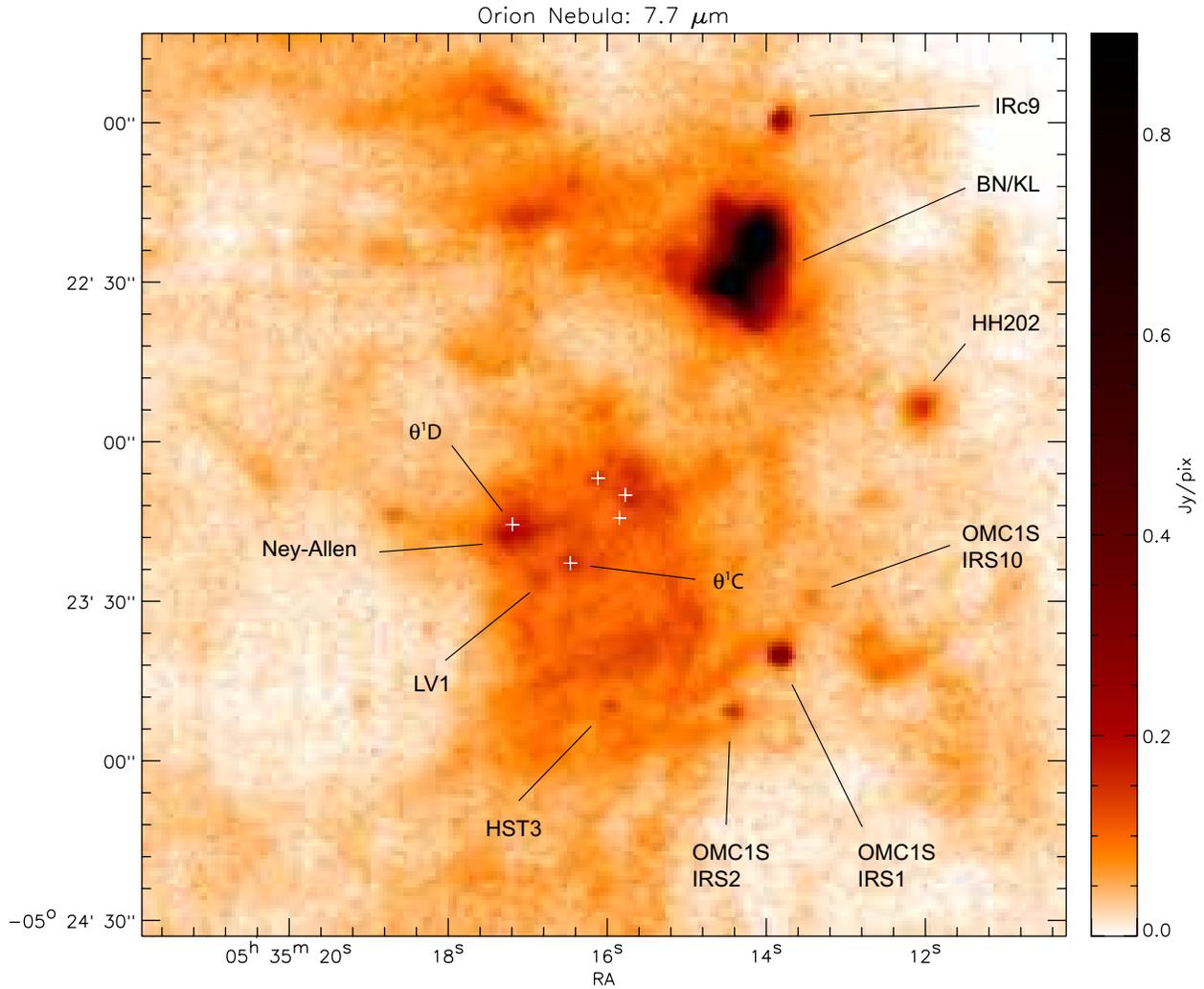}
\caption{FORCAST images of the Orion Nebula at 7.7~\micron, including prominent known sources. Details for the BN/KL region (saturated in this image) at the upper right are presented in a separate paper~\citep{DeBuizer:2012}.  White crosses mark the locations of the Trapezium stars ($\theta^1$ Ori).}
\label{fig:image_77}
\end{figure}

\begin{figure}
\epsscale{1.0}
\plotone{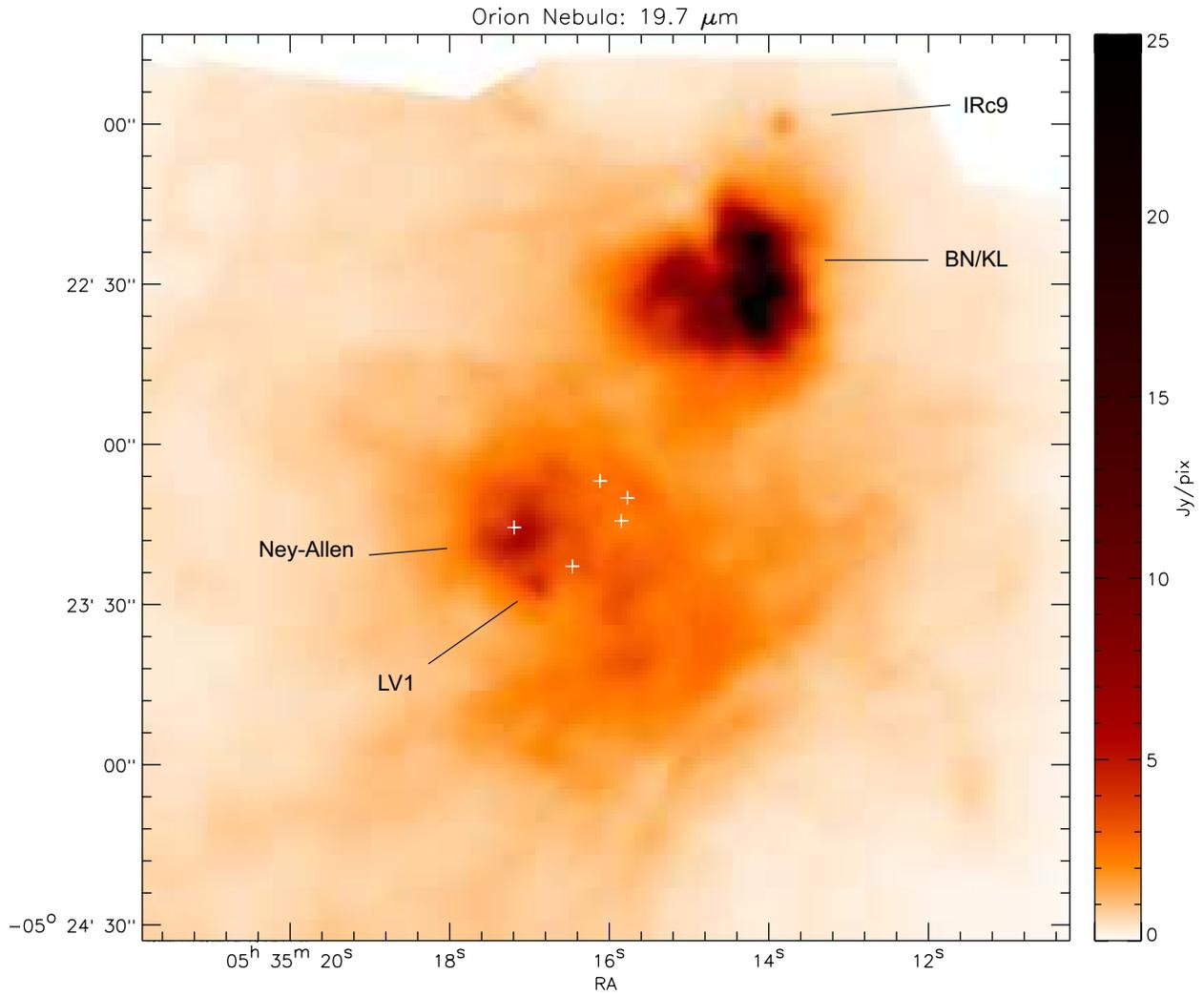}
\caption{FORCAST image of the Orion Nebula at 19.7~\micron.}
\label{fig:image_19}
\end{figure}

\begin{figure}
\epsscale{1.0}
\plotone{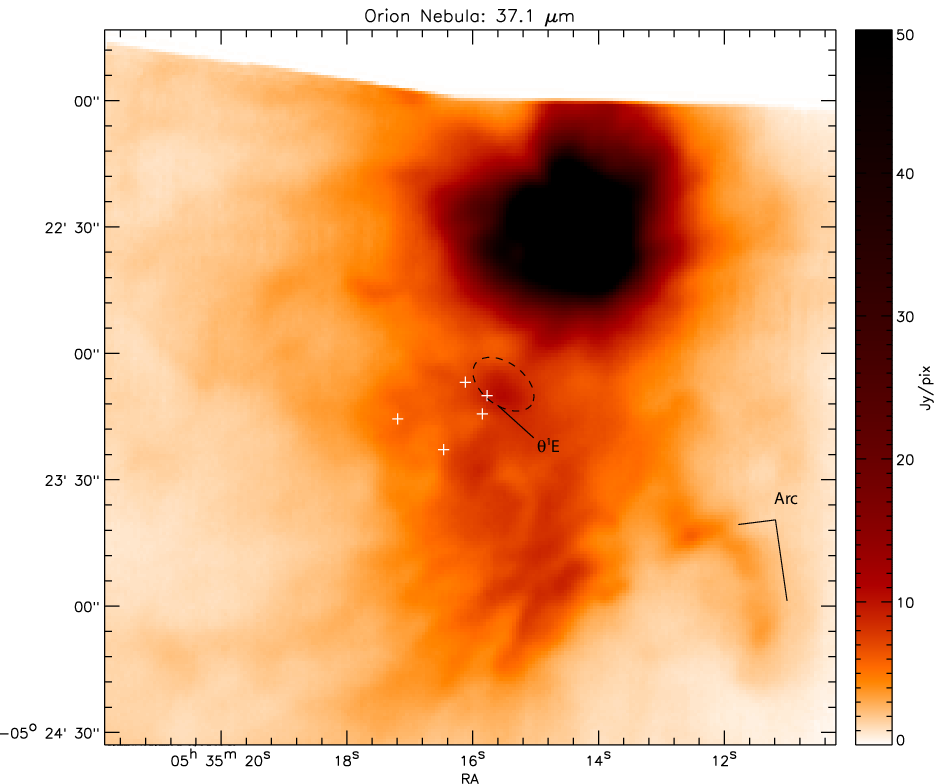}
\caption{FORCAST image of the Orion Nebula at 37.1~\micron.  The ``Arc'' of emission SW of the Trapezium and the bright feature NW of $\theta^{1}$~E~Ori (dashed oval) are discussed in Section~\ref{sect:diffuse}.  }
\label{fig:image_37}
\end{figure}

The FORCAST images were calibrated to a flux density per pixel using a standard instrument
response derived from measurements of standard stars
and solar system objects over several short science and basic science
flights~\citep{Herter:2012}. The color correction from a flat spectrum source to that of
the Orion sources is $\lesssim 2\%$. The estimated $3\sigma$ uncertainty in the
calibration due to variations in flat field, water vapor burden, and altitude
is approximately $\pm 20\%$. Flux calibrated images for the 7.7, 19.7, and 37.1~\micron\ data are shown in Figs.~\ref{fig:image_77}, \ref{fig:image_19}, and \ref{fig:image_37}:  the images at 6.6~\micron\ suffered from poor image quality and are not shown.

\begin{figure}
\epsscale{1.0}
\plottwo{fig4a.eps}{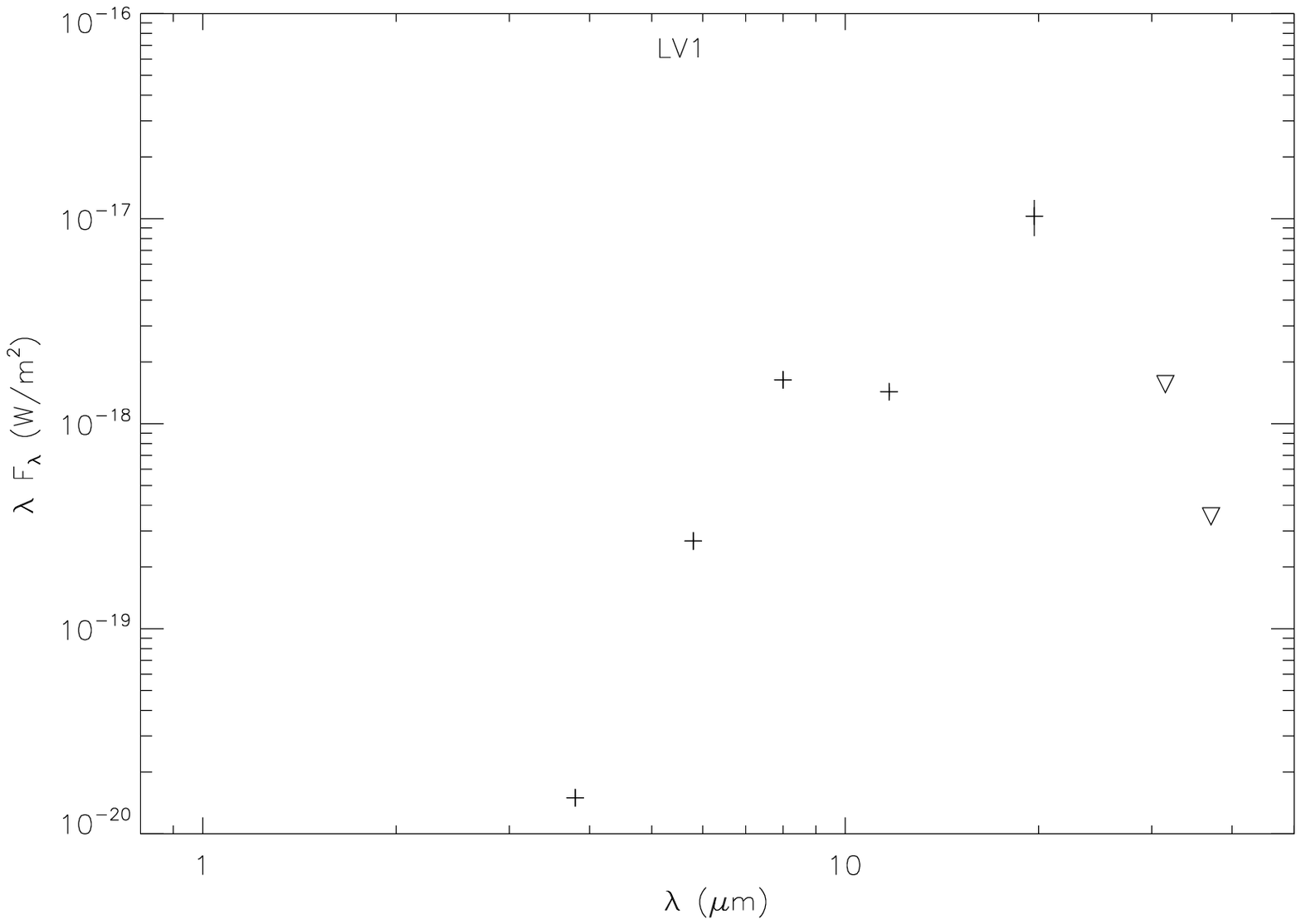}
\caption{
SEDs for IRc9 and  LV1.  
{\it Left Panel}---SED for IRc9, including our mid-IR FORCAST measurements, near-IR photometry from \citet{Muench:2002}, Spitzer IRAC fluxes~\citep{Megeath:2012}, and additional mid-IR ground-based measurements~\citep{Smith:2005}.  Photometry for IRc9\ at 19.7 and 31.5~\micron\ was done using a 5.4\arcsec\ aperture; note that IRc9\ was not observed at 37.1~\micron.  The best model fit from \citet{Robitaille:2007} for IRc9 is shown as the black line in the left panel.  The dashed line shows the stellar photosphere corresponding to the central source of the best fitting model, as it would look in the absence of circumstellar dust (but including interstellar extinction).  
{\it Right Panel}---SED for LV1, including our mid-IR FORCAST measurements, Spitzer IRAC fluxes~\citep{Megeath:2012}, and additional mid-IR ground-based measurements~\citep{Smith:2005}.  Photometry for LV1 at 19.7~\micron\ was done using a 3.8\arcsec\ aperture.  Upper limits (triangles) are 3-$\sigma$ and based on RMS noise in the background at the position of the source.
}
\label{fig:seds}
\end{figure}

Our FORCAST images at 7.7~\micron\ reveal a number of compact sources which are consistent with IRAC images of the Orion Nebula available from the Spitzer Legacy Archive and coincide with a number of known objects in the region (see Fig.~\ref{fig:image_77}). Of the known sources detected at 7.7~\micron, only \lvone\ and \ircnine\ are visible at 19.7~\micron, and no known compact sources are detected at 31.5 and 37.1~\micron.  Source fluxes (or upper limits) measured from our FORCAST images at 19.7, 31.5, and 37.1~\micron\ for \ircnine\ and \lvone\ are included in the SEDs shown in Figure~\ref{fig:seds}.  

%

\begin{figure}
\epsscale{1.0}
\plotone{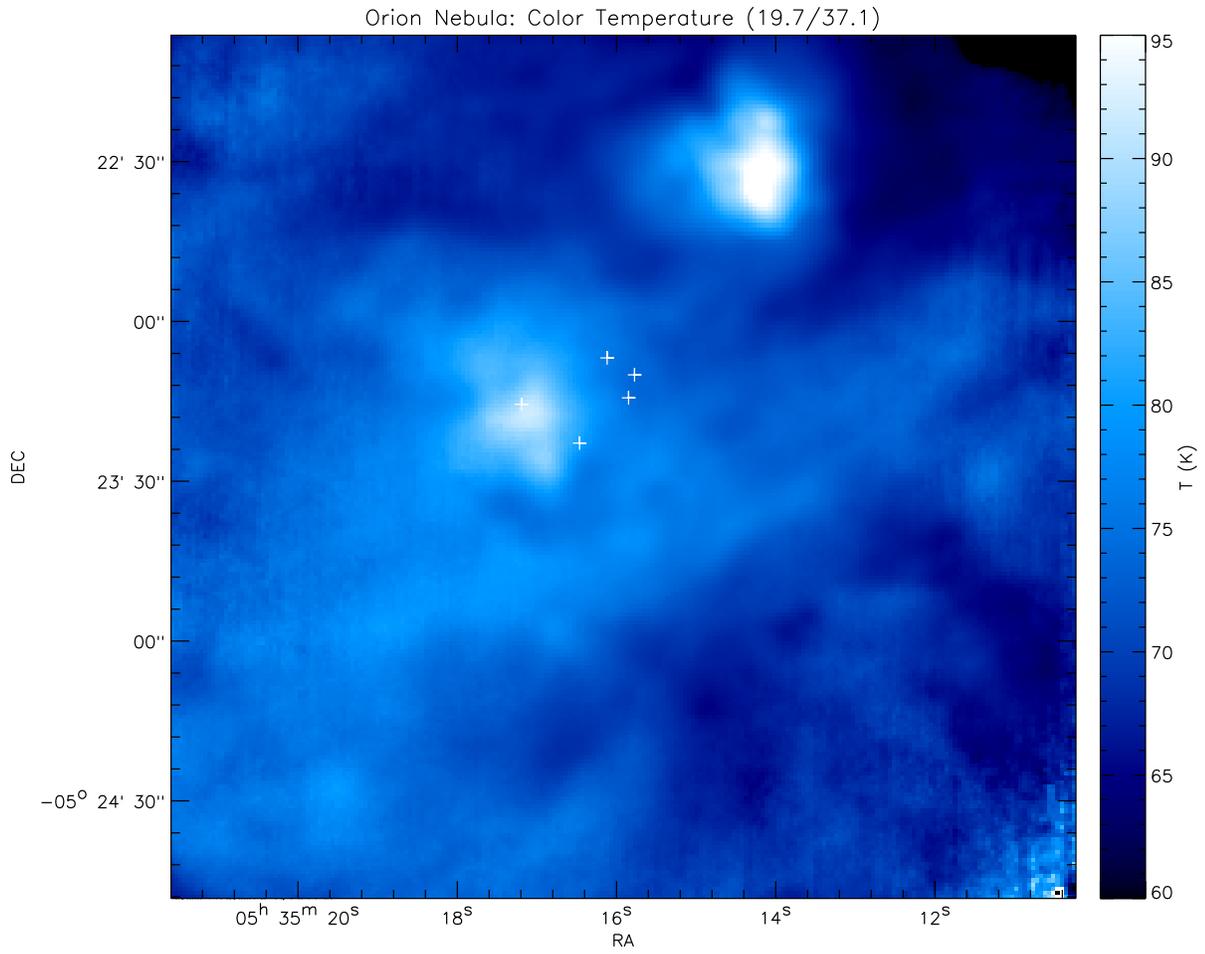}
\caption{Temperature map of the Orion Nebula from 19.7 and 37.1~\micron\ data.  White crosses mark the locations of the Trapezium stars ($\theta^1$ Ori).}
\label{fig:temp}
\end{figure}

A temperature map of the region was generated from the 19.7 and 37.1~\micron\ images using a least-squares procedure to fit the observed surface brightnesses in each pixel to the model
\begin{equation}
I_{\nu} = (1-e^{-\tau_{ref} (\tau/\tau_{ref})}) \times B_{\nu}(T_c)
\end{equation}
where $B$ is the blackbody function.   We have adopted the Mathis extinction law for the wavelength dependence of $\tau$ ($= \tau/\tau_{ref}$) and $\tau_{ref}$ is evaluated for 19.7~\micron\ (Fig.~\ref{fig:temp}). The surface brightnesses at each wavelength were given equal weight. The model corresponds to the solution of the radiative transfer equation for emission/absorption of material along the line of sight.  The median dust temperature throughout the central part of the nebula is $\approx$70K and the emission is optically thin.  The \object{Ney-Allen Nebula} is the warmest region in the nebula (besides BN/KL), with a color temperature in excess of $\approx$90K (see discussion in Section~\ref{sect:neyallen}).  These temperatures are significantly colder than those reported by \citet{Hayward:1994} for both the warm ($T\sim300$~K) and cold ($T\sim110$~K) dust components.  This is to be expected, however, since our study utilizes longer wavelengths than \citet{Hayward:1994} and hence is more sensitive to colder dust.

\section{Discussion}

\subsection{Ney-Allen Nebula}
\label{sect:neyallen}

The dominant feature of the Trapezium region is the \object{Ney-Allen Nebula}, which is bright in all the FORCAST filters except for 37~\micron, where it is very difficult to distinguish from variations in the background emission (Fig.~\ref{fig:image_37}).  The morphology of Ney-Allen is clearly wavelength-dependent in our FORCAST images: 
At 7.7~\micron, it is an elongated source (approximately 4\arcsec\ $\times$ 7\arcsec),  offset by 2.2 arcsec from \thetaoned\ (see Fig~\ref{fig:image_77});
in the images at 19 and 31~\micron, it is a crescent surrounding the position of \thetaoned\ with the apex pointing roughly toward a PA of 220\arcdeg. 
The crescent shape is also apparent at N and Q bands~\citep{Robberto:2005}, and 11.7~\micron ~\citep{Smith:2005}. In addition, the diameter of the crescent increases with wavelength, from approximately 5.5\arcsec\ at 19~\micron\ to 8.2\arcsec\ at 31~\micron, suggesting a radial temperature gradient indicative of heating by the central source, \thetaoned.   The same increase in size with wavelength (exceeding the increase in PSF size as a function of wavelength) is also evident in IRAC images of the Trapezium available from the Spitzer Legacy Archive.  

\citet{Robberto:2005} suggested that the Ney-Allen emission could be due to dust that was evaporated from a massive circumstellar disk associated with \thetaoned.  \citet{Smith:2005} argue against this interpretation citing lack of evidence for such a disk (though it is possible that the disk is now completely dissipated).  Since the axis of symmetry of the Ney-Allen crescent aligns very closely with the direction of proper motion for \thetaoned\ (217\arcdeg), \citet{Smith:2005} instead postulated that the Ney-Allen emission is due to dust swept up in the bow shock between the stellar wind of \thetaoned\ -- a B0.5V star~\citep{Mason:1998} -- and the ambient nebular material. Our SOFIA-FORCAST observations are consistent with this interpretation.  Because \thetaoned\ is the only member of the Trapezium to show such a compressed shell, then, if this scenario is correct, it is probably in a relatively high density portion of the overall nebula.  This suggests that \thetaoned\  is much closer to the background dense cloud than are the other members of the Trapezium, and may even be plunging into the background PDR~\citep{Smith:2005}.

\subsection{Proplyds}
\label{sect:proplyds}

Two of the compact sources detected in the 7.7~\micron\ images (\object[OW94 159-350]{HST3} and \lvone) are known proplyds---externally illuminated, low-mass young stellar objects whose circumstellar disks are being photoablated by the strong EUV and FUV radiation from \thetaonec ~\citep{Odell:1993,Bally+98}.   \lvone\ is also visible in the 19.7 and 31.5~\micron\ images, but is indistinguishable from the nebular background at 37.1~\micron.  At 19.7~\micron\ \lvone\ appears slightly elongated in the NE--SW direction (Fig.~\ref{fig:image_19}); the same orientation as the wind-wind bow shock (the stand-off shock between the proplyd evaporative flow and the stellar wind from \thetaonec) that is seen in both HST [O~III] and H$\alpha$ images~\citep{Bally+98} and at 11.7~\micron~\citep{Smith:2005}.  This suggests that dust from the proplyd circumstellar disk is entrained in the evaporative flow and then ``piles up'' at the wind-wind shock where it is heated by the UV radiation from \thetaonec\ and emits strongly in the mid-IR.  Both HST3 and \lvone\ are detected in 3.6~cm radio continuum~\citep{Zapata:2004}; only HST3 is detected in the COUP X-Ray survey~\citep{Getman:2005}.

We have combined our mid-IR flux measurements with IRAC fluxes~\citep{Megeath:2012}, and additional ground-based mid-IR measurements~\citep{Smith:2005} to generate the mid-IR SED for \lvone\ shown in Fig.~\ref{fig:seds}.  Interpretation of this SED is difficult, however, for two reasons:
\begin{enumerate}
\item \lvone\ is a known double~\citep[and references therein]{Graham:2002}, so mid-IR dust emission can arise from a circumstellar disk around either source and from the observed interaction zone between the two sources;
\item mid-IR emission from the wind-wind shock cannot be spatially separated from the binary proplyd emission.  
\end{enumerate}

Models of externally illuminated disks constructed by \citet{Robberto:2002} show that the mid-IR emission from 10 to 40~\micron\ is dominated by  emission from the disk interior (peaking at $\sim40$~\micron) and emission from the superheated atmosphere (peaking at $\sim25$~\micron).  These models tend to peak around 30 -- 60~\micron, which is not observed for \lvone\ (the peak in the \lvone\ SED appears to be around 20~\micron), suggesting that the mid-IR emission is dominated by a combination of the superheated disk atmosphere(s) of the two \lvone\ components and the dust piled up in the wind bow shock.  \lvone\ also has a modest 10~\micron\ silicate emission feature~\citep{Shuping:2006}, which is also a strong component of the superheated disk atmosphere~\citep{Robberto:2002}.

\subsection{Embedded Young Stellar Objects}
\label{sect:irc9}

Roughly 30\arcsec\ north of the BN/KL region is the relatively isolated infrared source \ircnine~\citep{Wynn-Williams:1974,Downes:1981}.  \ircnine\ is thought to be an embedded, luminous young stellar object (YSO) with a massive circumstellar disk~\citep{Smith:2005a}:  The near- to mid-IR SED shown in Fig.~\ref{fig:seds} is consistent with this interpretation.  
Model SEDs of young stellar objects were fit to the SED of \ircnine\ using the Online YSO SED Fitting tool~\citep{Robitaille:2007} with best fit shown in Fig.~\ref{fig:seds}.  The \citet{Robitaille:2007} models assume a single central stellar source with different combinations of axisymmetric circumstellar disks, infalling flattened envelopes and outflow cavities. The fitting procedure interpolates the model fluxes to the apertures used in the measurements, scaling them to a given distance range of a source, which in this case is taken to be 416$\pm6$~pc~\citep{Kim:2008}, and the line-of-sight extinction $A_V$.  We adopted a range of $A_V$ from 2.0 to 10.0, based on the OMC1 extinction map~\citep{Scandariato:2011}.  While these models yield estimates of physical parameters which are highly degenerate in general, they are fairly good at estimating the total luminosity of sources, which is our primary interest.  

The best fit SED models for \ircnine\ have a total bolometric luminosity of 110 to 215~\lsun, which is somewhat higher than the value found by \citet{Smith:2005a}, but agrees with their conclusion that \ircnine\ is the progenitor of an early-A (or late-B) ZAMS star with $M=$~3---4~\msun.  The best fit models also indicate that the bulk of the emission longward of 10~\micron\ is due to a large circumstellar envelope ($R> 2700$~AU), whereas the flux from 3---10~\micron\ is primarily from a much smaller circumstellar disk ($R< 100$~AU).  This suggests that the spatially resolved 8.8 to 18.3~\micron\ emission observed by \citet{Smith:2005a} and attributed to a large circumstellar disk, may in fact be the inner regions ($R<1000$~AU) of an envelope.

In addition to \ircnine\, we also detect three of the infrared sources in the OMC1-S region (\object[SBSMH 1]{IRS1}, \object[SBSMH 2]{IRS2}, \object[SBSMH 10]{IRS10}), also thought to be YSOs (Fig.~\ref{fig:image_77}).  At longer wavelengths, detection and flux measurement of the OMC1-S sources is not clear due to confusion with variations in the background  emission.  Limiting fluxes based on variations in the nebular background, however, suggest that 20 -- 40~\micron\ emission from the  OMC1-S sources may be significant.  We compared the 19, 31, and 37~\micron\ limiting fluxes to  flux measurements from Spitzer~\citep{Megeath:2012} and found that the SEDs from 4.5 through 31~\micron\ are consistent with Class~I or ``flat spectrum'' protostars, though it is not possible to rule out the possibility that they are highly-reddened Class~II sources.  The bright 7.7~\micron\ source SW of the BN/KL region coincides with \object{HH202}, a shock-excited region thought to be associated with an outflow or jet from OMC1-S \object[SBSMH 1]{IRS1}~\citep{Smith:2004}.  In contrast, \object{HH202} is very weak at 6.6~\micron\ and not detected in the 19, 31, and 37~\micron\ images at all, strongly suggesting that the 7.7~\micron\  excess is due to PAH emission.


\subsection{Diffuse Emission}
\label{sect:diffuse}

The diffuse emission observed in our FORCAST images is generally a combination of emission from  dust in both the background MIF and the \ion{H}{2} region itself.  The observed 19.7/37.1~\micron\ color temperature values across the nebula (Fig.~\ref{fig:temp}) are consistent with cool dust in the background PDR.  The structures apparent in our 19.7 and 37.1~\micron\ images are consistent with higher-resolution ground-based Q-band (20~\micron) observations obtained by \citet{Robberto:2005}.  The filamentary structure SW of the Trapezium coincides roughly with bright [N~II] and \halpha\ emission features~\citep{ODell:2000} and also with 8.4 and 1.5 GHz radio continuum emission~\citep{Dicker:2009}.  \citet{Robberto:2005} concluded that the filaments in the mid-IR correspond to undulations or folds in the background PDR surface that cause limb-brightening and shadowing effects---similar to the Orion Bar but on a smaller scale.  


The diffuse emission drops significantly in the OMC1S region in all of our FORCAST images.  The 19.7/37.1 color temperature drops significantly as well, indicating reduced UV heating in this region.  We interpret this as a shadowing effect wherein the bright folds and undulations NE of OMC1S correspond to dense material that blocks the UV radiation field from \thetaonec.  This is consistent with the interpretation of OMC1S as a cold, dense region of gas and dust embedded {\em within} the \ion{H}{2} region~\citep{ODell:2009}.  We suggest that the formation of the embedded protostars in OMC1S may have been triggered by  advancing shock waves when the Orion Nebula was first forming.   

The 37.1~\micron\ image presented here is the highest spatial resolution image of the Orion Nebula at this wavelength to date.  There are a number of features in the 37.1~\micron\ image that have no counterparts at shorter wavelengths.  Most prominent, is a $\sim$60\arcsec\ arc of emission to the west of OMC1S (see Fig.~\ref{fig:image_37}).   This feature is not evident in narrow-band emission~\citep{ODell:2000}, but is present in the 1.5 GHz radio continuum~\citep{Dicker:2009}.  The arc coincides with a region of low color temperature as well (Fig.~\ref{fig:temp}).  Interpretation of this arc is unclear, but it seems plausible that it is heated internally by the embedded OMC1S protostars.  

Most of the other structures observed at 37~\micron\ coincide with bright regions of [NII] and \halpha\ emission, seen both in natural~\citep{Bally+98} and extinction-corrected images~\citep{ODell:2000}.  In particular, there is a very bright extended source just NW of the Trapezium (also noted by \citet{Wynn-Williams:1984} in their low-resolution 30~\micron\ image) which corresponds to one of the brightest regions in the nebula in \halpha\ and [\ion{N}{2}] emission~\citep{ODell:2000}.  Some of the diffuse structures at 37.1~\micron\ are also coincident with fluctuations in the color temperature (Fig.~\ref{fig:temp}).  These two phenomena suggest that these are all  perturbations in the surface of the background PDR.  One possible exception, however, is the emission WSW of \thetaonec, which coincides with roughly 5 known proplyds, including the prominent sources LV4, 5, and 6~\citep{Bally+98}.  We suggest that this is emission from dust that is entrained in the photoablation flows off the circumstellar disks around these sources and then heated by the UV radiation field of \thetaonec.

\acknowledgements

We thank R.~Grashius, S.~Adams, H.~Jakob, A.~Reinacher, and U.~Lampater for their SOFIA telescope engineering and operations support. We also thank the SOFIA flight crews and mission operations team (A.~Meyer, N.~McKown, C.~Kaminski) for their SOFIA flight planning and flight support. Based on observations made with the FORCAST instrument on the NASA/DLR Stratospheric Observatory for Infrared Astronomy (SOFIA).  SOFIA science mission operations are conducted jointly by the Universities Space Research Association, Inc. (USRA), under NASA contract NAS2-97001, and the Deutsches SOFIA Institut (DSI) under DLR contract 50 OK 0901.  Financial support for RYS was provided by NASA through SSI award 01-830.0 issued by USRA.  Financial support for TH and JA was provided by NASA through award 8500-98-014 issued by USRA for the development of the FORCAST instrument.

{\it Facilities:} \facility{SOFIA (FORCAST)}


\bibliographystyle{../apj}




\end{document}